**Theoretical and experimental investigation of optical absorption anisotropy in β-Ga$_2$O$_3$**

F. Ricci[1], F. Boschi[2], A. Baraldi[2], A. Filippetti[3], M. Higashiwaki[4], A. Kuramata[5], V. Fiorentini[1], and R. Fornari[2]

[1] *Department of Physics, University of Cagliari, Cittadella Universitaria, 09042 Monserrato (CA), Italy*

[2] *Department of Physics and Earth Sciences, University of Parma, Area delle Scienze 7/A, 43124 Parma, Italy*

[3] *CNR-IOM, UOS Cagliari, Cittadella Universitaria, 09042 Monserrato (CA), Italy*

[4] *National Institute of Information and Communication Technology, 4-2-1 Nukui-Kitamachi, Koganei, Tokyo 184-8795, Japan*

[5] *Tamura Corporation, 2-3-1, Hirosedai, Sayama-shi, Saitama, 350-1328, Japan*

*ABSTRACT*

The question of optical bandgap anisotropy in the monoclinic semiconductor β-Ga$_2$O$_3$ was revisited by combining accurate optical absorption measurements with theoretical analysis, performed using different advanced computation methods. As expected, the bandgap edge of bulk β-Ga$_2$O$_3$ was found to be a function of light polarization and crystal orientation, with the lowest onset occurring at polarization in the ***ac*** crystal plane around 4.5-4.6 eV; polarization along ***b*** unambiguously shifts the onset up by 0.2 eV. The theoretical analysis clearly indicates that the shift of the ***b*** onset is due to a suppression of the transition matrix elements of the three top valence bands at Γ point.

*INTRODUCTION*

The semiconducting sesquioxides, and among them β-Ga$_2$O$_3$, have been employed for several decades as transparent conducting oxide (TCO) electrodes for fabrication of solar cells, displays, electronic, and opto-electronic devices [1-3]. The interest was confined to such application as the conductivity of these metal-oxides was invariably n-type, and attempts to get an effective p-type doping failed. The key requirements of TCO electrodes are indeed high electrical conductivity and good transparency, while crystallographic perfection is a minor issue. Furthermore, for a long period no high-quality substrates and epilayers were available, which in turn impeded the development of a truly full-oxide electronics. Recently, β-Ga$_2$O$_3$ has attracted renewed attention, as large single crystals [4-6] and high-quality homo- and hetero-epitaxial layers became available [7-9], which paved the way to novel application areas, namely: substrates for GaN-based LEDs [10, 11] and high-power transistors [12, 13]. As in many previous cases, this technological breakthrough triggered much research on the fundamental materials properties, which in turn produced improved materials.

Some basic properties of the stable β phase of Ga$_2$O$_3$ are known with good degree of confidence. Its crystallographic structure, for example, belongs to the monoclinic system (space group C2/m) with lattice constants *a* = 12.23 Å, *b* = 3.04 Å, *c* = 5.80 Å, and an angle γ=103.73° between *a* and *c* [1, 2, 14]. With regard to optical and electronic properties, first-principles calculations and Angular-Resolved Photo-Electron Spectroscopy (ARPES) [15-18] concur that the minimum gap is between the Γ point of the conduction band and the (quasi) M point of the valence band. However, being the valence band quite flat, the difference between the valence band maxima at M and Γ reduces to a few meV. Therefore, the effective near-edge optical behaviour is, essentially, that of a Γ direct-bandgap material, with largely anisotropic valence bands and (practically) isotropic conduction band. The band gap values of bulk β-Ga$_2$O$_3$, measured by optical absorption, generally fall in the range 4.5 to 4.9 eV; see for example [6, 19-22]. This scattering derives from the anisotropy of optical bandgap first reported in the pioneering work of Matsumoto et al. [21] in 1974, followed by a reappraisal in 1997 by Ueda et al. [22]. Using UV light polarised along the directions *b* and *c* of (100) Ga$_2$O$_3$ platelets, these authors showed that the bandgap values were higher by about 0.15-0.27 eV when **E**∥*b*. A good agreement exists for the **E**∥*c* values (4.50 and 4.52 eV in [21] and [22], respectively), whilst for **E**∥*b* the agreement between the two works was somewhat qualitative (4.65 [21] and 4.79 eV [22]). The reasons of this discrepancy was not discussed.

In spite of the bandgap anisotropy being a recognised experimental fact, it is surprising that many recent papers reported bandgap values without providing the crucial information about sample orientation and measurement geometry. The principal motivation to the present work was indeed to shed more light on the optical properties - band structure relationships in bulk β-Ga$_2$O$_3$. Accurate optical absorption experiments were carried out on high-quality melt-grown gallium oxide substrates, in order to verify if and how the conclusions of [21] and [22] apply to different materials and crystal orientations. Experimental data were also obtained for the orientations (-201) and (010), not investigated so far. The bandgap values are critically discussed in view of the results of *ab initio* computation and a physical interpretation of the optical anisotropy is provided.

*EXPERIMENTAL AND THORETICAL METHODS*

The samples investigated in this work were (010) and (-201)-oriented wafers cut from boules grown by Edge-defined Film Fed Growth at Tamura Corporation. They were nominally undoped (carrier concentrations 2.4 and 1.7 10$^{17}$ cm$^{-3}$ respectively) and 0.65 mm thick. Polarized absorption was measured at room temperature and normal incidence in the 800-200 nm range by means of a Varian 2390 spectrophotometer and a Glan-Taylor polarizer. As described further below, we oriented the electric field vector **E** parallel to the crystallographic axes ***a***, ***b***, and ***c***. It must be noted that what is denoted for simplicity **E**||***a*** is actually the direction orthogonal to the ***bc*** plane, which indeed is 13° away from ***a***. Additional measurements were carried out on Sn-doped samples (1.0-3.2x10$^{18}$ cm$^{-3}$) and it was found that the absorption edges well match those of the undoped counterparts, when the same set-up geometry is employed. At this level of doping the spectra of doped and undoped samples only differed in the infrared region but not in the near bandgap region. Therefore, the following discussion on bandgap anisotropy will essentially focus on the results obtained on undoped β-Ga$_2$O$_3$.

To interpret and support the experimental results, *ab initio* calculations using several density-functional-related methods were performed. Specifically, hybrid (HSE) [23] and self-interaction corrected (VPSIC) [24, 25], GGA (generalized gradient) and LDA (local density) functionals, and many-body perturbation theory in the non-self-consistent G$_0$W$_0$ and self-consistent-Green's-function GW$_0$ [26] approximations were used. HSE and GW are the VASP-PAW implementations [27-30]. The level of agreement with experimental data was generally good, always well within 10% and down to 2% on the absolute values of the gap. Confirming recent results on magnetic Mott insulators [32] the best performance in comparison with experiment is that of VPSIC. The technical ingredient were essentially standard [33].

*RESULTS AND DISCUSSION*

Figure 1 reports the spectra obtained in the case of nominally undoped β-Ga$_2$O$_3$ at normal incidence on the (010) surface with polarizations **E**‖*c* and **E**‖*a*. Correspondingly, absorption edges at 4.54 and 4.57 eV were observed. The absorption anisotropy between **E**‖*c* and **E**‖*a* is thus limited to 0.03 eV.

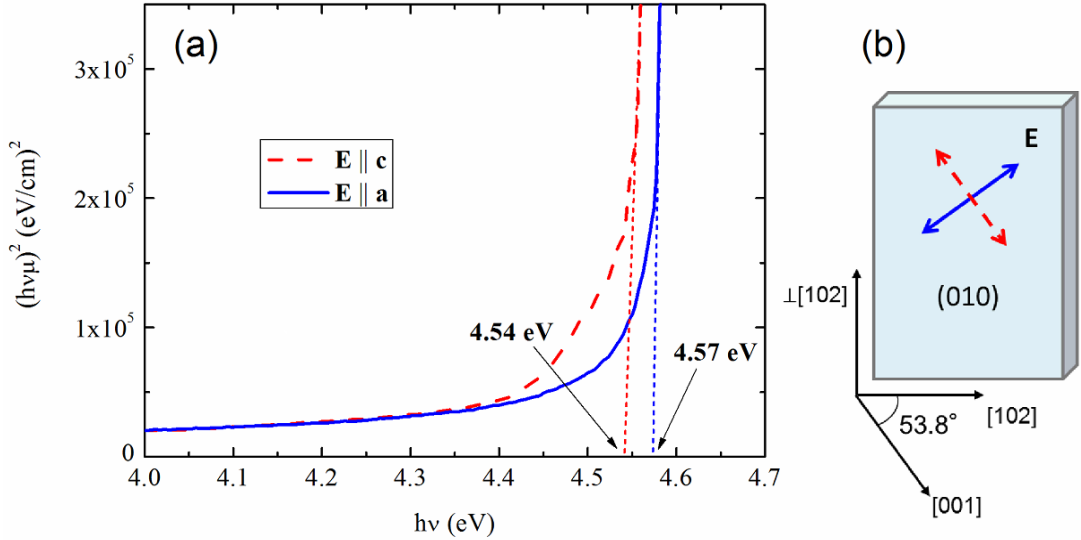

FIG. 1. *Absorption spectra of a (010) wafer at normal incidence for polarization **E**‖c and **E**‖a, respectively (a), and schematic of sample orientation, axes, and polarizations (b).*

In Figure 2 we report the absorption spectra obtained on a (-201) oriented wafer. With this sample geometry, it is easy to probe the **E**‖*b* configuration as well as **E**‖[102] (clearly *b*⊥[102]). The **E**‖[102] absorption edge at 4.52 eV is in line with those just reported for *c* and *(*quasi*) a* polarization, as expected from [102] being a combination of the *a* and *c* vectors. However, the onset of the absorption for **E**‖*b* is at 4.72 eV, that means 0.2 eV higher in energy. This significant shift of the absorption edge is a direct evidence for anisotropy and confirms the previous observations in [21, 22].

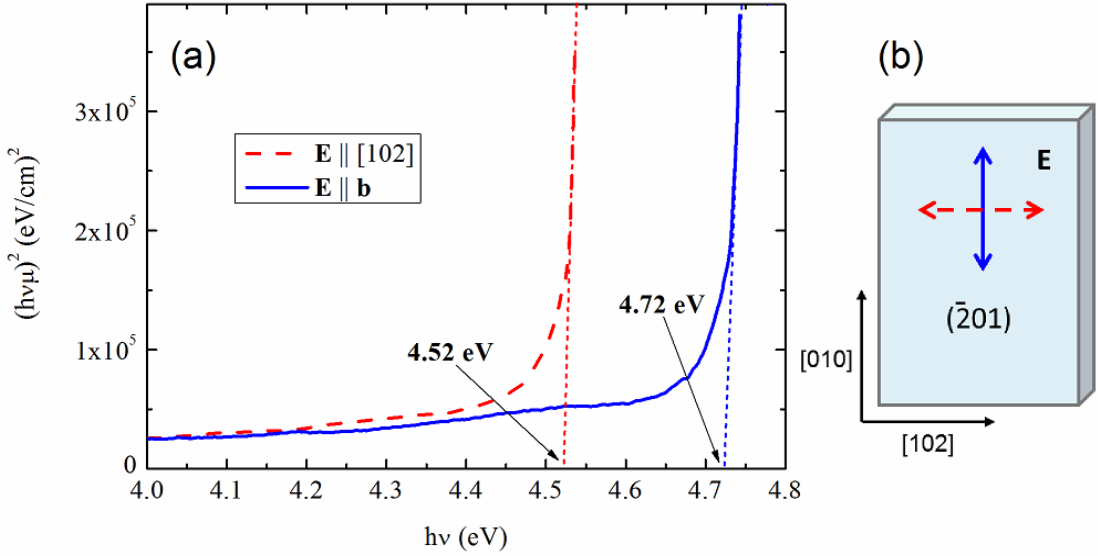

FIG. 2. *Absorption spectra of a (-201) wafer with polarization E∥b and E∥[102], respectively (a), and schematic of sample orientation, axes and polarization (b).*

Summarizing the experimental results for polarisation parallel to the three crystallographic axes, one can see that the lowest absorption edge at 4.54 eV is for polarization **E**∥*c*, followed by the close 4.57 eV onset when **E**∥*a*, while **E**∥*b* gives a clearly separated 4.72 eV.

In Figure 3 we report the absorption coefficient μ predicted by VPSIC-GGA at the experimental volume and internal coordinates relaxed according to quantum forces. As for the Figs. 1 and 2, the absorption coefficient is plotted in Tauc form with exponent 2, so that it linearizes for a direct dipole-allowed gap (the direct gap at Γ is indeed almost identical to the indirect minimum gap, which is invisible in absorption on the intensity scale of the direct transition). The edge of the absorption is obtained by linear extrapolation of $(h\nu\mu)^2$ to zero absorption. μ is obtained in the standard way from the calculated imaginary part of the frequency-dependent dielectric tensor. The latter is almost diagonal, but sizably anisotropic due to the low symmetry of the crystal.

Clearly, there are distinct absorption edges as function of polarization, and their energetic order is **E**∥*c* < **E**∥*a* < **E**∥*b*, in agreement with experiments. For **E**∥*c* the onset is at 4.65 eV compared to experimental 4.54 eV: 2% off is an excellent result for the standards of ab initio theory, which may miss gap values by as much as 100%. For **E**∥*a* the onset is higher by just 0.1 eV, which compares well with the 0.03 eV shift observed experimentally (see Fig. 1). Again a very satisfactory result, also considering the measurement was not exactly matching the *a* axis.

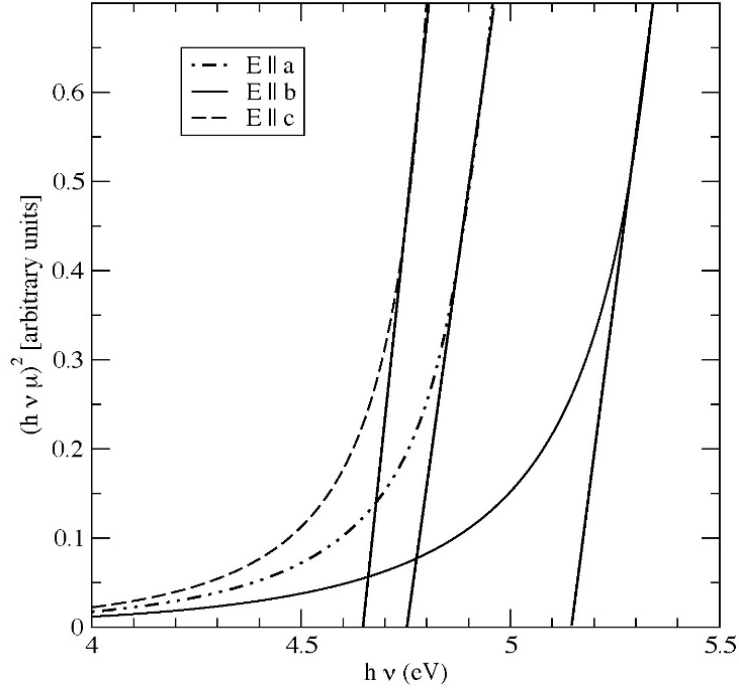

FIG. 3. *Tauc plot of the absorption coefficient, showing the polarization-dependent onsets.*

For **E**‖***b***, the theoretical onset is 0.5 eV higher than for **E**‖***c***, again in satisfactory agreement with experimental data (Fig. 2). Although the VPSIC theoretical results are able to reproduce well the relative sequence and separation of bandgap edges, we have to note that experimental values are systematically lower than theoretical predictions. One possible explanation for this systematic error might be the high absorbance of wafers, which ultimately can lead to a slight underestimation of the bandgap edges.

It is worth recalling that even within corrected DFT methods, gap errors are commonly in the order of a few percent and, as often as not, gaps can be overestimated (for PSIC see e.g. [25], Fig.4); thus, theory-experiment deviations such as those we find are not unexpected. It is also worth mentioning that, as usual in the business of *ab initio* optical properties, we implicitly identify eigenvalues and eigenvectors of Kohn-Sham equations as quasiparticle energies and states. This is justified by the Kohn-Sham equations being formally identical to Hedin-Lundqvist quasiparticle equations if the self-energy is identified with the exchange-correlation potential or its corrected versions [25,32].

In Figure 4 we reported the direct gap obtained by all the different computation methods as a function of the crystal cell volume. Some considerations have to be made when looking at the data presented in this graph. First, theoretically predicting the crystal cell volumes has always some inherent uncertainty (in this case VPSIC provides the volume closest to experiment). Secondly, the dependence on microscopic structural details is modest (VPSIC eigenvalues in the experimental and GGA cells are almost identical). Finally, the pressure derivative of the gap extrapolated from the data is 3 meV/kbar, in line with previous reports [34]. The agreement with experiment is overall reasonable. Partially self-consistent $GW_0$ is closest to experiment and VPSIC, which is in itself an interesting theoretical outcome.

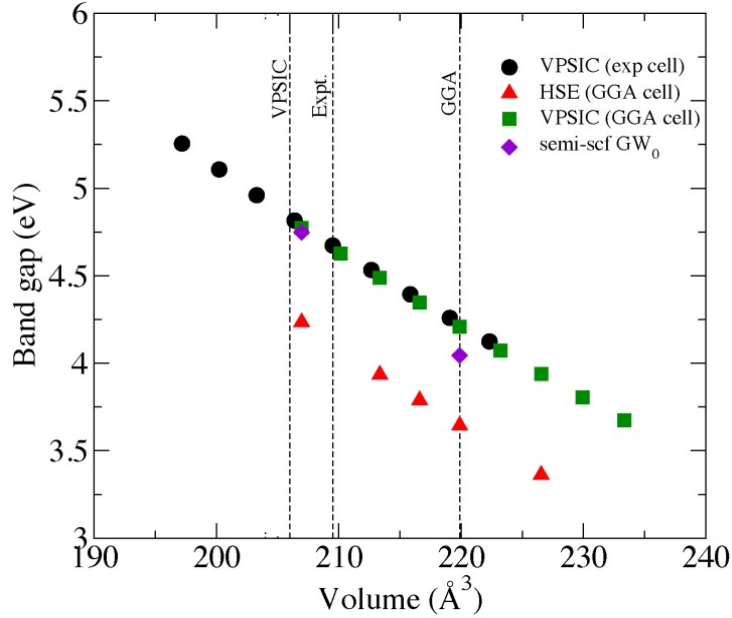

FIG. 4. *Gap vs. volume as obtained from hybrid and self-interaction corrected functionals, as well as from GW0 many-body perturbation theory. The line Exp marks the cell volume obtained from experimental lattice parameters [1].*

As a final point, we observe that the upwards shift of the onset for **E**∥**b** stems from a suppression of the matrix elements for the transition from the top valence bands. The absorption is proportional to $|\mathbf{e}\cdot\mathbf{P}_{vc}|^2$, where **e** is the polarization and $\mathbf{P}_{vc}=(\langle P_a\rangle, \langle P_b\rangle, \langle P_c\rangle)$, where e.g. $\langle P_b\rangle$ is the matrix element between valence and conduction states of the momentum component $P_b$ along the **b** axis. A smaller $\langle P_b\rangle$ implies weaker absorption for **E**∥**b**.

In Figure 5a we show the matrix elements along the three axes at the Γ point, along with relevant bands and wavefunctions. Clearly $\langle P_b\rangle$ from the top three valence bands is essentially zero, and this causes the suppression of the absorption. The reason is that the initial and final wavefunctions sketched in Fig. 5b have the same parity along **b**, and therefore the matrix-element integral is zero [35] (in addition, the valence states have a very small amplitude). The recovery of the matrix element at the fourth-from-top band below the valence top shows that the anisotropy shift is essentially the distance of that specific band from the valence top.

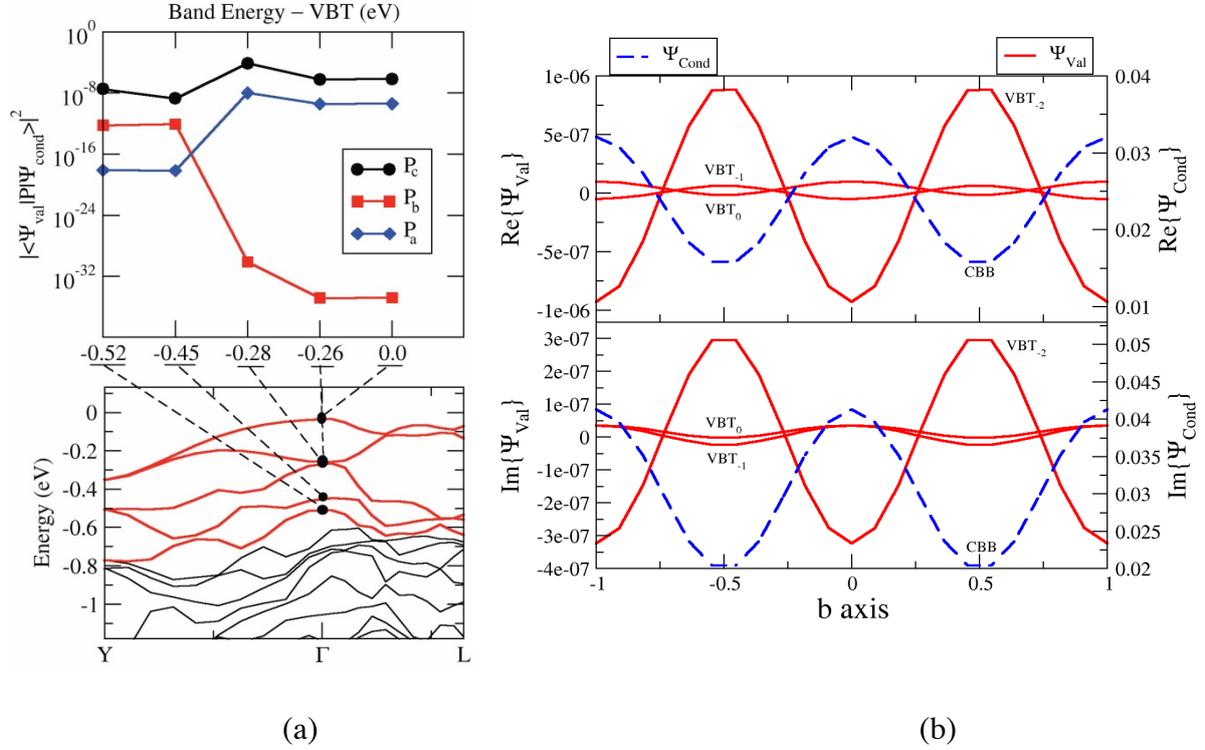

(a) (b)

FIG. 5. *(a) Matrix element (top) along the crystal axes vs energy from valence top, and bands near the valence top. Along **b** the matrix element is suppressed (note the log scale). The bands show that the anisotropy shift of the **E**//**b** onset is related to the distance of the top to fourth-from-top bands. (b) Along **b** the Γ-point wavefunction Ψ(n,y; x=z=k=0) of the three top valence bands are very small, and have the same parity as the conduction band (real part: top; imaginary part: bottom; note the different scales).*

## CONCLUSIONS

In conclusion, the optical absorption of β-$Ga_2O_3$ was seen to be anisotropic as function of incident light polarization and crystal orientation, in that confirming previous reports. In this work, the experimental measurements were extended to unexplored wafer orientations. The lowest onset was at about 4.55 eV in (010)-oriented wafers. In this case, whatever the polarization on the *ac* plane, only minor shifts of the bandgap were detected. For (-201)-oriented samples, light polarization along *b* has the strongest effect as it shifts the absorption edge towards higher energy by 0.2 eV. Theoretical analysis (especially by the VPSIC method) supports these observations and provides a physical interpretation of the absorption anisotropy. In the specific case of the higher absorption edge for **E**∥***b***, theory indicates that the higher energy is actually due to suppression of the transition matrix elements of the three top valence bands.

Acknowledgments

Work partly supported by MIUR PRIN 2010 'Oxide', Fondazione Banco di Sardegna, CAR of Uni Cagliari, and CINECA. The PhD grant of F.B. is supported by Fondazione Cariparma.